# Lithography-free patterning of chalcogenide materials for integrated photonic devices

Authors: Zhen Hu[1]†, Yuru Li[1]†, Yan Li[1]*, Shunyu Yao[1]*, Hongfei Chen[1], Tao Zhang[1], Zhaohuan Ao[1], Zhaohui Li[1,2]*

*Correspondence: Yan Li (liyan329@mail.sysu.edu.cn); Shunyu Yao (yaoshy23@mail.sysu.edu.cn); Zhaohui Li (lzhh88@mail.sysu.edu.cn)

[1]Guangdong Provincial Key Laboratory of Optoelectronic Information Processing Chips and Systems, School of Electrical and Information Technology, Sun Yat-sen University, Guangzhou 510275, China
[2]Southern Marine Science and Engineering Guangdong Laboratory (Zhuhai), Zhuhai 519000, China

†These authors contributed equally to this work.

**Abstract**

Chalcogenide material-based integrated photonic devices have garnered widespread attention due to their unique wideband transparency. Despite their recognized CMOS compatibility, the fabrication of these devices relies predominantly on lithography techniques. However, chalcogenide thin films are highly susceptible to oxidation, necessitating customized process flows and complex protective measures during lithography. These requirements are hardly compatible with current commercial CMOS manufacturing platforms designed for silicon photonics, significantly limiting the practical applications of chalcogenide photonic devices. In this work, we ingeniously exploit the ease of oxidation of chalcogenide materials, presenting a novel laser-induced localized oxidation technique for spatial patterning on chalcogenide thin films, enabling concise lithography-free fabrication of chalcogenide integrated photonic devices. Using $Sb_2S_3$ as an example, we experimentally demonstrate localized multi-level oxidation with a sizable overall refractive index contrast of 0.7 at near-infrared, featuring a high spatial resolution of 0.6 μm. Based on this technique, multiple integrated photonic devices are demonstrated, showing versatile functionalities, including color printing at visible and metasurface-based spatial light modulation at near-infrared regions. Leveraging the inherent phase-change property of $Sb_2S_3$, an active Fresnel zone plate, enabling switchable beam focusing, is further demonstrated, indicating the feasibility of concise fabrication of active photonic devices. Our work offers a brand-new modulation dimension for chalcogenide materials and provides a significantly simplified approach for realizing chalcogenide-integrated photonic devices.

**Introduction**

With the rapid advancement of information technology, global communication data is growing exponentially, imposing higher demands on information processing chips in communication, storage, and optical computing[1-4]. Photonic integrated devices, which leverage the high bandwidth and parallelism of optical information processing, have become a crucial area of development[5-8]. Recently, alongside traditional silicon photonic devices, new materials such as III-V semiconductors, lithium niobate, and chalcogenide glass have emerged, offering promising opportunities for integrated optical signal generation and modulation[9-11]. Chalcogenide glass has been widely used in infrared imaging due to its unique broadband transparency[12-14]. With advancements in nanofabrication techniques, chalcogenide materials are now being further utilized in integrated photonic devices[15-18]. Their high nonlinearity and low-threshold phase-change properties provide exciting possibilities, particularly for optical modulation and detection[19,20]. In recent years, chalcogenide integrated photonic devices have emerged as a focal point of research in photonics. The rapid advancement is largely attributed to the unique broadband transparency of chalcogenide materials, which has opened up vast possibilities for designing photonic devices across the near-infrared to mid-infrared spectrum[20-23]. Researchers have successfully developed on-chip chalcogenide resonators with ultra-high quality factors, which have optimized the performance of integrated optical transmission and paved the way for highly sensitive on-chip sensing techniques[24,25]. The high nonlinear refractive index characteristic of chalcogenide materials has further propelled the study of nonlinear optics[26]. Coupled with the materials' intrinsic ultra-low loss properties, various high-performance all-optical signal processing technologies have been demonstrated, including parametric frequency conversion, on-chip optical frequency comb generation, and on-chip optical amplification[15,27]. Moreover, the non-volatile phase-change characteristics of chalcogenide materials have introduced innovative approaches for both on-chip and spatial light modulation, indicating significant potential for applications in optical computing and optical data storage[28].

Despite the extensive research, the widespread application of chalcogenide photonic devices still faces numerous challenges. One of the key bottlenecks is the stable fabrication of high-quality chalcogenide material thin films and integrated photonic devices. Current studies generally believe that chalcogenide materials are compatible with standard CMOS processes[29,30]. However, these materials inherently have a low damage threshold and are highly susceptible to oxidation, which complicates the lithography process of chalcogenide photonic devices[31]. In recent years, considerable research has been dedicated to fabricating high-quality chalcogenide material thin films and photonic devices, yielding substantial outcomes. By constructing a trapezoidal waveguide structure combined with high-quality chalcogenide material thin films, researchers have achieved a chalcogenide resonant microcavity with an ultra-high quality factor of 1.44×10$^7$ using arsenic sulfide (AsS) material[32]. Regarding the fabrication of general-purpose photonic devices, previous studies have continuously refined the etching process of chalcogenide materials, achieving the fabrication of chalcogenide waveguides with ultra-low loss in the near-infrared range[33-35]. However, it is important to note that the reported fabrication processes are highly customized and difficult to directly integrate with the standard CMOS processing platforms developed by the industry primarily for silicon-based photonic devices. Therefore, an urgent need remains to advance new integrated photonic device processing technologies adaptable to chalcogenide materials, offering high flexibility and low complexity.

In this study, we introduce a versatile patterning technique for chalcogenide thin films, which facilitates a streamlined, lithography-free fabrication process for chalcogenide integrated photonic devices. Capitalizing on the inherent susceptibility of chalcogenide materials to oxidation[36,37], we have developed a novel approach that leverages laser-induced localized oxidation. The thermal effect of the laser significantly enhances and accelerates the oxidation process, leading to pronounced modulation of the refractive index. Employing a high numerical aperture laser for precise focusing, our method achieves spatial resolution nearing the diffraction limit, offering a compelling alternative for fabricating chalcogenide integrated photonic devices. As a case in point, we demonstrate multilevel oxidation patterning on antimony sulfide ($Sb_2S_3$) thin films, achieving an overall refractive index modulation depth exceeding 0.7, which substantially surpasses the changes attainable with traditional photo-induced methods. Integrating this with high-precision laser direct writing technology, we have attained a spatial resolution of approximately 0.6 μm, enabling the effective creation of chalcogenide metasurfaces in the near-infrared spectrum. This advancement underscores the considerable potential of our technique in applications such as imaging and communication[38,39]. Furthermore, the intrinsic phase-change properties of $Sb_2S_3$ allow for dynamic modulation of the optical response of the fabricated metasurfaces, broadening the scope of our technique to include active photonic devices. Our approach stands out from traditional lithography techniques by effectively circumventing the material damage associated with exposure and etching processes. Moreover, our method significantly reduces the fabrication complexity by enabling single-step processing of photonic devices. This innovation is of significant importance for propelling the application and development of chalcogenide integrated photonic devices, marking a pivotal step forward in the field.

**Results**

**Laser-induced oxidation patterning on chalcogenide thin film**

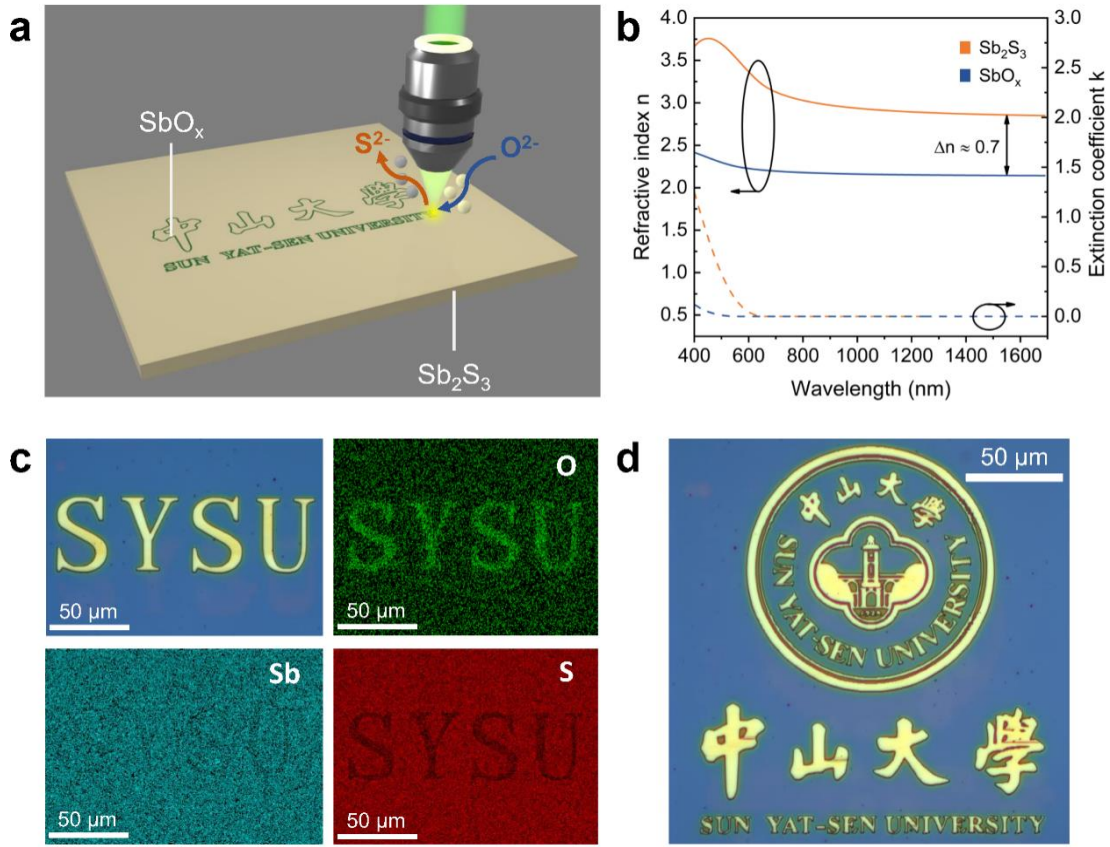

**Figure 1. Laser-induced oxidation patterning on $Sb_2S_3$ thin film.** a. Schematic of the laser-induced localized oxidation technique indicating the working principle. b. Measured reflective index and extinction coefficient spectrum of $Sb_2S_3$ material before and after oxidation patterning. c. The optical microscope image of the patterned "SYSU" letters on a thin $Sb_2S_3$ film and the characterized element distribution of oxygen, antimony, and sulfur show effective oxidation of the patterned area. d. The optical image of the printed emblem of Sun Yat-sen University indicates the capability of fabricating complex patterns.

Figure 1a illustrates the schematic of refractive index patterning on chalcogenide thin film based on laser-induced localized oxidation. Here, $Sb_2S_3$ thin film is employed as a representative material for demonstration. Under ambient air conditions, a high numerical aperture (NA) objective lens focuses a laser beam onto the $Sb_2S_3$ thin film. The thermal effect of the laser induces rapid oxidation of the irradiated $Sb_2S_3$ material, transforming it into antimony oxide ($SbO_x$), offering substantial refractive index contrast, which is pivotal for the fabrication of integrated photonic devices. Our experimental characterization details the variations of the refractive index of the $Sb_2S_3$ thin film across the visible and near-infrared spectrum before and after laser-induced oxidation. As depicted in Figure 1b, the oxidation process achieves a refractive index modulation exceeding 0.7 across both the visible and near-infrared bands, indicating wideband feasibility. Notably, the material retains its broadband transparency post-oxidation, with the extinction coefficient remaining near zero beyond 550 nm, a feature crucial for realizing high-quality integrated photonic devices. Coupled with a high-precision displacement platform, our technique enables flexible refractive index patterning on the chalcogenide thin film, meeting the fabrication requirements for a variety of photonic devices (for detailed information on the laser processing platform, please refer to Section S1 of Supplementary Information). Utilizing a commercialized laser centered at 532 nm in conjunction with an objective lens featuring NA = 0.95, we have achieved a high spatial resolution of nearly 0.6 μm, closely approaching the diffraction limit (for detailed characterization of the spatial resolution; please refer to the Section S2 of Supplementary Information). With high-precision grating scanning technology, we have demonstrated effective exposure of complex patterns on a thin $Sb_2S_3$ film with a thickness of 100 nm on top of a silicon substrate. Figure 1c showcases the printing of the "SYSU" logo using our laser processing technique. The characterized element distribution indicates the enrichment of oxygen and the depletion of sulfur at patterned areas, with the antimony remaining approximately unchanged, confirming the realization of laser-induced oxidation of chalcogenide materials. The significant refractive index contrast induced by material oxidation can markedly alter the visible spectrum of the processed area, enabling efficient nano-printing. From the microscopic images given in Figure 1c, it can be observed that the laser-processed areas exhibit uniform color differences. Further characterization of the $Sb_2S_3$ thin film before and after laser processing using atomic force microscopy (AFM) reveals that the oxidized regions maintain a low surface roughness, with an average roughness of only 2.30 nm. This further demonstrates the significant application potential of the current patterning technique in the visible light spectrum (for detailed AFM characterization results, please refer to Section S3 of Supplementary Information). Building on this technology, we further demonstrate nano printing of the Sun Yat-sen University emblem, confirming the capability to process complex patterns. The characterized element distribution in Section S4 of Supplementary Information indicates effective localized oxidation.

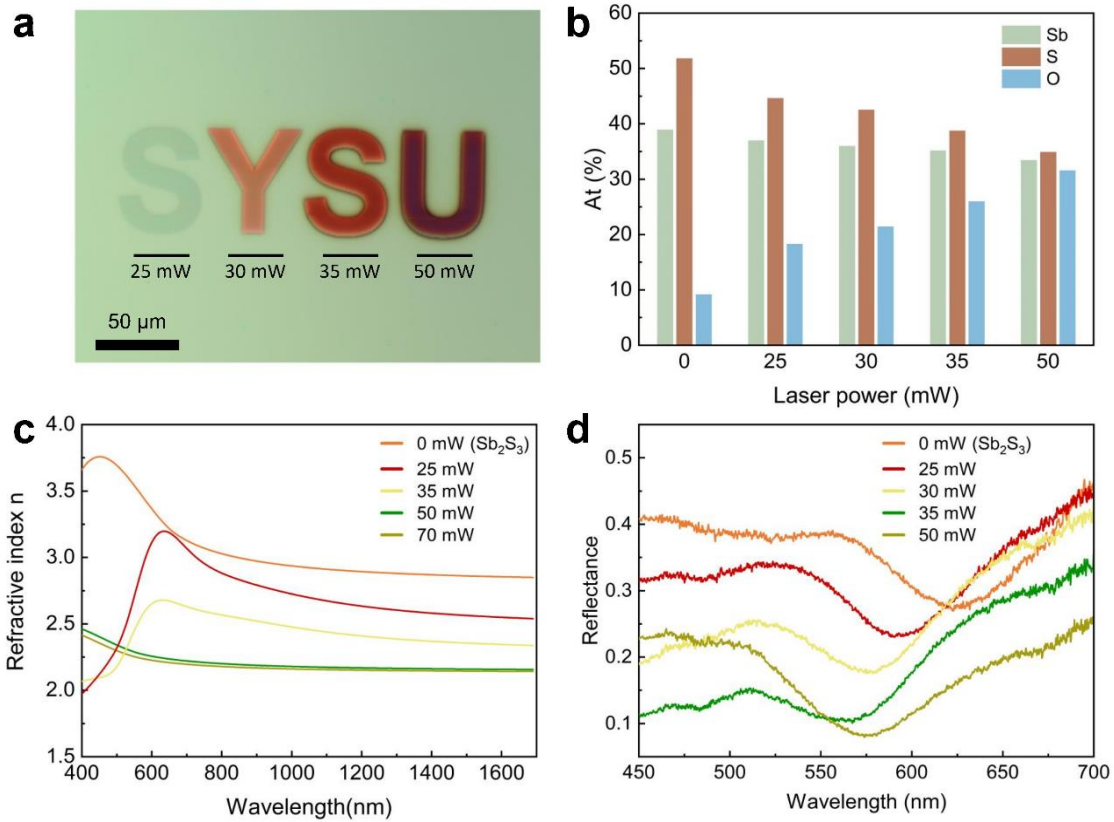

**Figure 2 Multi-stage controllable laser-induced oxidation.** a. Photograph of the color-printed patterns, enabled by tuning the laser power during processing. b. Characterized proportion of oxygen, sulfur, and antimony in the patterned areas using different laser power. c. Measured refractive index at visible and near-infrared regions patterned with varying laser power. d. Measured reflectance spectrum of color printed areas at visible range.

Notably, the oxidation of chalcogenide materials is an intrinsic continuous process. Consequently, by modulating the laser processing parameters, quasi-continuous control over the degree of oxidation in the irradiated area can be achieved, thereby enabling multi-stage refractive index modulation. In this study, by adjusting the power of the excitation laser, we have demonstrated four levels of oxidation control based on $Sb_2S_3$ thin films, providing a new dimension in the design of chalcogenide photonic devices.

Figure 2a presents microscopic images of the oxidized regions in $Sb_2S_3$ thin films induced by laser at different excitation powers. Distinct color differences can be observed under various laser powers. Here, we compare the ratio of oxygen, sulfur, and antimony elements between different processing areas. As shown in Figure 2b, with the increase in laser power, the proportion of oxygen in the exposed area gradually increases, as expected. In contrast, the proportion of sulfur gradually decreases, and the proportion of antimony remains unchanged essentially, fully verifying the multi-stage controllable oxidation degree of the processing area. The significant color differences between different regions indicate that as the degree of oxidation changes, the material's refractive index also experiences noticeable differences, making multi-stage exposure of chalcogenide-integrated photonic devices possible. Characterization of the refractive index in different processing areas clearly shows, as depicted in Figure 2c, that with the increase in excitation power, the degree of oxidation of the chalcogenide material in the exposed area leads to a gradual decrease in the effective refractive index in the near-infrared band, eventually stabilizing. There is a nice linear correspondence between the degree of material oxidation and the modulation of the effective refractive index. Therefore, the integrated photonic device processing method based on laser-induced localized oxidation has excellent potential for achieving effects similar to grayscale exposure. By finely controlling the excitation power, we anticipate that multi-stage quasi-continuous refractive index modulation can be realized in the near-infrared band, which is of significant importance for the flexible fabrication of integrated photonic devices. The refractive index modulation caused by laser-induced oxidation is more complex in the visible light band. From the refractive index spectrum given in Figure 2c, it can be observed that under different degrees of oxidation, in addition to the refractive index, the dispersion of chalcogenide materials in the visible light band also changes, further expanding the potential application scenarios of the current processing method. Taking micro-nano printing as an example, by combining the multi-stage oxidation exposure method with the design of chalcogenide thin film, color printing, as shown in Figure 2a, can be achieved in the visible light band. Here, a 200 nm $Sb_2S_3$ thin film on a silicon substrate is an example for demonstration. By characterizing the reflection spectrum of different processing areas, it can be seen that the $Sb_2S_3$ thin film has an absorption peak near 620 nm in its initial state, making the initial state appear nearly green. With the increase in the degree of laser-induced oxidation, the absorption peak of the exposed area in the visible light band continuously shifts towards the shorter wavelengths, making the area color gradually move towards a red hue, thereby achieving effective color printing.

**Lithography-free fabrication of chalcogenide metasurfaces**

As mentioned above, color printing exemplifies the fundamental capabilities of the laser-induced localized oxidation exposure technique. By employing a short-wavelength laser centered at 532 nm in combination with a high NA objective lens, our technique attains a spatial resolution of approximately 0.6 μm, nearing the diffraction limit of the laser used for processing. This resolution is deemed adequate for the precision required to fabricate most integrated photonic devices, especially for the near-infrared band.

Inspired by the metal-insulator-dielectric (MID) metasurface design paradigm, our work incorporates a multilayer stack of an $Sb_2S_3$ thin film atop a reflective gold layer, separated by a thin aluminum oxide dielectric insulator. We create a two-dimensional periodic dot array by inducing localized oxidation on the $Sb_2S_3$ surface. The oxidation process reduces the effective refractive index in the exposed areas, establishing a contrast with the unexposed regions and effectively emulating two-dimensional micro-hole arrays. Fine-tuning the oxidation level and period of the dot arrays allows effective supporting of hybrid plasmonic resonances in the near-infrared, facilitating additional amplitude and phase modulation. Figure 3a displays a microscopic image of the metasurface fabricated via the laser-induced localized oxidation method. Figure 3b provides detailed fabrication insights into the two-dimensional dot arrays. The high-precision electric displacement platform ensures a uniform spatial distribution of the dot arrays, which is critical for producing high-quality plasmonic resonance modes. The refractive index variation due to material oxidation in the exposed area is visually distinct, presenting a noticeable color contrast with the unexposed area. The reflection spectrum characterization in the near-infrared band confirms the anticipated effective resonance excitation. The metal's intrinsic absorption at the resonance central wavelength significantly reduces metasurface reflectivity, achieving effective reflection modulation. As depicted in Figure 3c, adjusting the period of the dot arrays allows for precise control over the resonance central wavelength, enabling targeted wavelength manipulation. The amplitude modulation capability of chalcogenide metasurfaces forms a fundamental building block for spatial light field control. Extending this concept, we construct a Fresnel zone plate, as shown in Figure 3d, validating the fabrication technique's applicability. A 0.95 μm period dot array is employed for strong absorption at 1550 nm, creating high-contrast amplitude modulation in conjunction with the strong reflection of the unexposed area. Referencing the classic design of a Fresnel zone plate, alternating ring-shaped exposure areas are prepared, as shown in Figure 3d. The Fresnel zone plate design details can be found in Figure S5 of Supplementary Information. Figure 3e shows the near-infrared imaging of the fabricated Fresnel zone plate at the target wavelength, confirming the expected alternating bright and dark ring distribution. Moreover, we characterized the evolution of the reflected light field of the fabricated Fresnel zone plate at the target wavelength as it propagates. As shown in Figure 3f, the beam tracing results verify effective beam focusing, with the measured focal length close to the designed value of 5 mm. A detailed methodological introduction of the beam tracing of the Fresnel zone plate can be found in Figure S6 of Supplementary Information.

It warrants emphasis that, while this study employs spatial light field amplitude modulation as an illustrative example, the chalcogenide materials retain their broadband low-loss properties, particularly in the near-infrared band post-laser oxidation. Coupled with sophisticated metasurface design, the current fabrication methodology is well-positioned to produce integrated photonic devices with high-efficiency phase modulation capabilities.

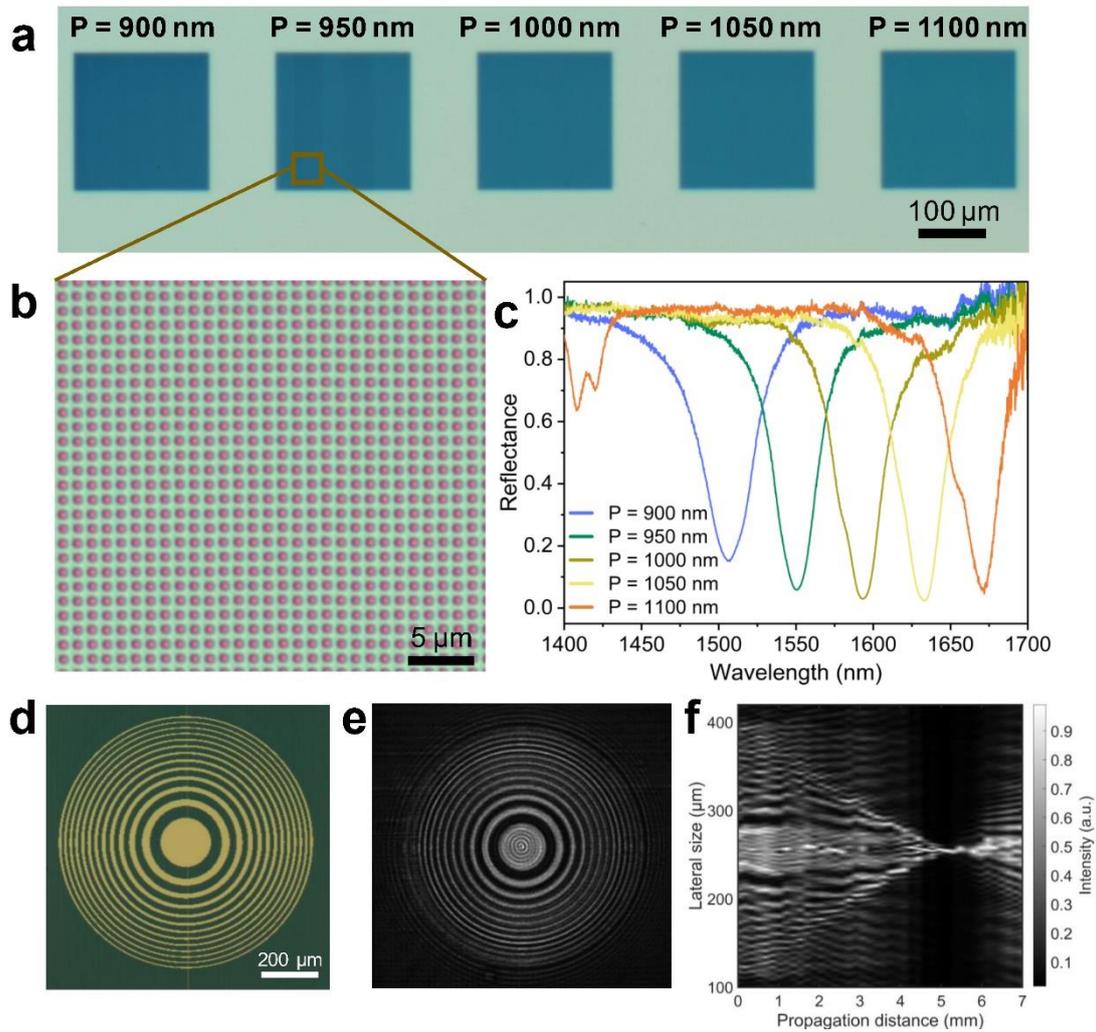

**Figure 3 Fabricated Fresnel zone plate at near-infrared based on laser-induced localized oxidation technique.** a, b. Optical microscope image of the fabricated two-dimensional dot arrays based on localized oxidation of $Sb_2S_3$ with varying periods. c. The characterized reflectance spectrum of fabricated dot arrays at near-infrared indicates effective resonance excitation. d. Optical microscope image of the fabricated Fresnel zone plate formed by alternating ring-shaped exposure areas. e. An optical image of the Fresnel zone plate at 1550 nm shows an alternating arrangement of bright and dark areas. f. Characterized beam propagation reflected by the Fresnel zone plate showing effective beam focusing with a focal length of nearly 5 mm.

**Reconfigurable oxidation-patterned chalcogenide phase change metasurfaces**

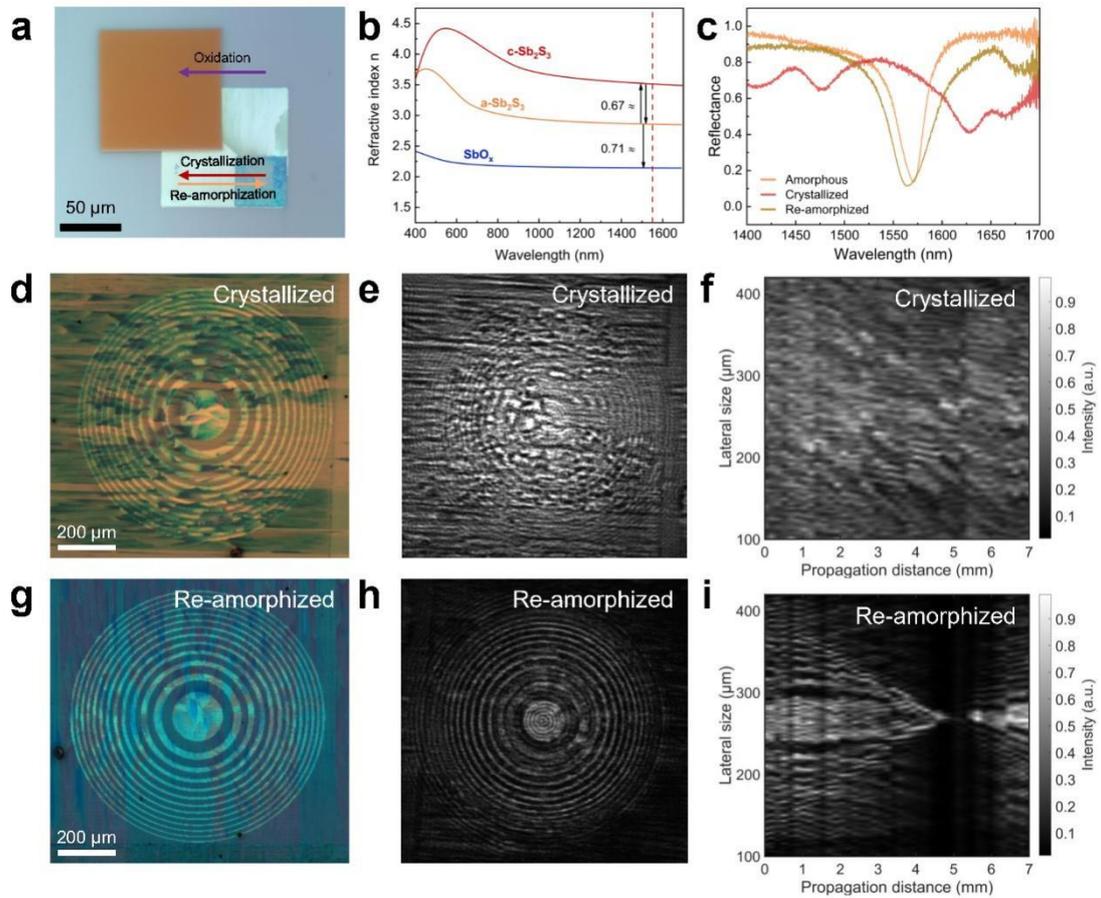

**Figure 4 Reconfigurable phase change modulation of an oxidation-patterned Fresnel zone plate.** a. Optical microscope image of the reconfigurable phase change modulation of the thin amorphous $Sb_2S_3$ film with a laser-induced pre-oxidation pattern. An $Al_2O_3$ layer caps the film, preventing the $Sb_2S_3$ from further oxidizing during the modulation. b. The measured refractive indices of the amorphous (a- $Sb_2S_3$), crystallized (c-$Sb_2S_3$), and oxidized ($SbO_x$) regions on the $Sb_2S_3$ film. c. The reflectance spectra of the amorphous, crystallized, and re-amorphized regions on the $Sb_2S_3$ film. d, e, f. Optical microscope, optical image at 1550 nm, and beam propagation determined the lost focus for the crystallized oxidation-patterned Fresnel zone plate, respectively. g, h, i. Optical microscope, optical image at 1550 nm, and beam propagation determined the focus again for the re-amorphized Fresnel zone plate.

Except for oxidation, $Sb_2S_3$ is also known as a mature phase change material. Under the excitation of external stimuli such as light and heat, the $Sb_2S_3$ material can freely transition between amorphous and crystalline states. It offers a substantial refractive index contrast, commonly used to realize active photonic devices[40]. In our work, the $Sb_2S_3$ material in the unexposed areas of the metasurfaces fabricated via laser-induced localized oxidation still retains its phase-change characteristics. Consequently, the optical response of the fabricated metasurfaces can also be dynamically modulated by altering the phase state of the material in the unexposed regions.

To verify this idea, phase transitions of the $Sb_2S_3$ film with a pre-oxidation pattern are performed. A continuous laser facilitates crystallization, while a pulsed laser induces re-amorphization. A dense layer of $Al_2O_3$ is covered atop the $Sb_2S_3$ film to prevent it from further oxidation and deformation during the phase transition. The reversible phase transition between its amorphous and crystalline states can be observed in Fig.4a. Meanwhile, the phase transitions are ineffective in the region where the $Sb_2S_3$ has been pre-oxidized. This means pre-oxidation patterns on amorphous $Sb_2S_3$ can maintain their optical functionality. Hence, the oxidation-patterned $Sb_2S_3$ metasurface is compatible with the phase change modulation. The significant contrast between the refractive indices of the amorphous and crystalline $Sb_2S_3$ as well as the oxidized regions (Fig.4b) indicates the inherent advantages of confining light by oxidation-patterned microstructures of $Sb_2S_3$ and dynamically tailoring optical responses by switching its phase state. The oxidation and the crystallization for an amorphous $Sb_2S_3$ result in opposite trends for tailoring the refractive index, providing a remarkable modulation depth with a maximum of ~1.38. It ensures great advantages for oxidation-patterned functional devices to enlarge their optical contrast under different states or broaden their operating bandwidth.

For example, the reflectance of the oxidation dot lattices as the Fresnel zone plate units in Fig.3d exhibits a large variation between the amorphous and crystalline states of $Sb_2S_3$. The unexposed $Sb_2S_3$ plays a role as the tunable dielectric environment (Fig. 4c). Meanwhile, a similar optical response is observed for the amorphous and re-amorphized states of $Sb_2S_3$, referring to its reconfigurable modulation. Figs. 4d to 4i demonstrate crystallization and re-amorphization modulation of the Fresnel zone plate. The outline of the Fresnel zone plate imaged by an optical microscope can be easily clarified for the crystallized and re-amorphized states (Figs. 4d and 4g). Meanwhile, under the illumination of a laser source operating at the designed optical band of the microlens and imaging by a CCD camera, only several ambiguous morphological details can be found in crystallized states (Figs. 4e). On the contrary, the feature of the Fresnel zone plate can be captured after re-amorphization (Fig. 4h), showing optical response characteristics different with the crystallized state. The switchable light focusing can be further verified using ray tracing along the light propagation direction. As shown in Figs. 4f and 4i, after crystallization, the two-dimensional dot arrays patterned through the oxidation process cease to modulate amplitude at the target wavelength, effectively transforming the Fresnel zone plate into a reflective surface. Upon re-amorphization, the optical response of the dot arrays is restored, endowing the zone plate with the capability to refocus light, with the focal length remaining at approximately 5 mm. This phenomenon underscores the tunable optical functionality of the fabricated Fresnel zone plates, demonstrating their potential for dynamic optical applications.

**Discussion**

This study introduces an innovative technique for refractive index patterning in chalcogenide materials through laser-induced localized oxidation, facilitating a lithography-free fabrication process for chalcogenide photonic devices. Our approach adeptly utilizes the innate oxidative susceptibility of chalcogenide materials, offering a highly precise and flexible method for device fabrication that circumvents the limitations of traditional lithography processes. Taking $Sb_2S_3$ as a representative chalcogenide material, this work demonstrates multi-stage controllable refractive index modulation based on laser oxidation, achieving a refractive index contrast exceeding 0.7 across the visible and near-infrared spectrum. By exploiting the substantial refractive index difference before and after material oxidation, coupled with a high-precision displacement platform, our work showcases the color printing of complex patterns on chalcogenide thin films, thereby validating the processing capabilities of the current patterning exposure technique. Notably, chalcogenide materials retain broadband low-loss characteristics after oxidation, which is important for fabricating near-infrared photonic devices. Utilizing a short-wavelength excitation laser with a high numerical aperture objective lens, we achieve spatial resolution close to 0.6 μm, providing a promising fabrication platform for near-infrared devices. Drawing on the classic MID metasurface design philosophy, we employ two-dimensional dot arrays created by laser-induced oxidation to fabricate chalcogenide Fresnel zone plates in the near-infrared band, demonstrating beam-focusing capabilities. Furthermore, leveraging the intrinsic phase-change properties of $Sb_2S_3$ material by controlling the phase state transition of unexposed areas, the fabricated Fresnel zone plates can achieve dynamically switchable focusing under external stimulation, thereby expanding the potential application scenarios of the current fabrication technique to active photonic devices. It should be noted that while this work primarily focuses on $Sb_2S_3$ material, the susceptibility to oxidation is a common characteristic of chalcogenide materials. We believe the current processing technique can readily extend to other chalcogenide materials. Additionally, the fabrication method relies on a platform similar to direct laser writing (DWL) technology designed for photoresists. Therefore, the current technique's spatial resolution and fabrication speed can be effectively enhanced by adopting advanced techniques common in existing DWL technology, such as beam shaping and parallel processing. In summary, our work opens new horizons for designing and applying chalcogenide-integrated photonic devices. The simplification of the fabrication process and the introduction of active modulation capabilities collectively contribute to advancing the field and hold promise for future innovations in photonic integrated circuits and systems.

## Materials and methods

### Sample preparation

The specimens depicted in Fig.1 and 2 were prepared by thermally evaporating $Sb_2S_3$ thin films of the respective thicknesses onto silicon substrates. Following laser-induced localized oxidation exposure, a 10 nm $Al_2O_3$ protective layer was deposited using atomic layer deposition (ALD) to prevent further oxidation of the samples. Specifically, the samples shown in Fig.1 utilized a 100 nm thick $Sb_2S_3$ layer, whereas the samples in Fig.2 employed a 200 nm thick $Sb_2S_3$ layer. The metasurfaces and Fresnel zone plates in Fig.3 and 4, also fabricated on silicon substrates, feature a stacked structure of a gold layer, an $Al_2O_3$ dielectric spacer, and an $Sb_2S_3$ thin film. Following substrate pretreatment, a 5 nm titanium (Ti) / 100 nm gold (Au) layer was deposited using electron beam evaporation to fabricate the metallic mirror. Subsequently, a 10 nm $Al_2O_3$ layer was applied via ALD to provide dielectric insulation. A 100 nm thick layer of $Sb_2S_3$ was thermally evaporated onto the sample. Similarly, after laser oxidation exposure, a 10 nm $Al_2O_3$ layer was deposited onto the sample surface to shield it from further environmental oxidation. Additionally, the protective aluminum oxide layer on the sample surface ensures that the oxidation-exposed areas are unaffected by the phase change excitation beam during the laser-induced phase transition process. In the current work, crystallization of the samples was achieved using a continuous-wave laser with a central wavelength of 532 nm. In comparison, re-amorphization was performed using a femtosecond laser with a pulse width of 350 fs and a central wavelength of 532 nm.

**Acknowledgments**

The project was supported by National Natural Science Foundation of China (NSFC) (62375291), Southern Marine Science and Engineering Guangdong Laboratory (Zhuhai) (SML2023SP231), the Joint Funds of the National Natural Science Foundation of China (Grant No. U23A20372) and the Program of Marine Economy Development Special Fund (Six Marine Industries) under Department of Natural Resources of Guangdong Province (Project No. GDNRC [2024]16).

**Author Details**

[1]Guangdong Provincial Key Laboratory of Optoelectronic Information Processing Chips and Systems, School of Electrical and Information Technology, Sun Yat-sen University, Guangzhou 510275, China. [2]Southern Marine Science and Engineering Guangdong Laboratory (Zhuhai), Zhuhai 519000, China

**Conflict of interest**

The authors declare no competing interests.